# Multi-channel all-optical switching based on coherent perfect absorption in atom-cavity system

Liyong Wang[1,*], Yinxue Zhao[2] and Jiajia Du[3,*]

[1]Department of Applied Physics, Wuhan University of Science and Technology, Wuhan 430081, China

[2]Wuhan Social Work Polytechnic, Wuhan 430079, China

[3]Chongqing University of Post and Telecommunications, Chongqing 400065, China

*Corresponding author: wangliyong@wust.edu.cn; dujj@cqupt.edu.cn

**Abstract.** We propose an ultrahigh-efficiency, broadband and multi-channel all-optical switching scheme based on broadband coherent perfect absorption (CPA) in a linear and nonlinear regimes in a cavity quantum electrodynamics (CQED) system. Two separate atomic transitions are excited simultaneously by two signal fields coupled from two ends of an optical cavity under the collective strong coupling condition. Three polariton eigenstates are produced which can be tuned freely by varying system parameters. The output field intensities of multiple channels are zero when the CPA criterion is satisfied. However, destructive quantum interference can be induced by a free-space weak control laser tuned to the multi-polariton excitation. As a consequence, the CQED system acts as a coherent perfect light absorber/transmitter as the control field is turned on/off the polariton resonances. In particular, the proposed scheme may be used to realize broadband multi-throw all-optical switching in the nonlinear excitation regime. The proposed scheme is useful for constructing all-optical routing, all-optical communication networks and various quantum logic elements.

Light is the basic carrier to transport and transfer information between different physical systems in fields of quantum information [1], quantum teleportation [2], all-optical computation [3, 4], etc. Compared to traditional electrical communications, all-optical communication have several advantages such as ultrahigh speed, extra-large bandwidth and superior security [5]. To construct various all-optical networks and perform the basic logic functions, it is an essential requirement to realize the high-efficiency and wide bandwidth light-controlling-light operations. However, this is still a great challenge today. All-optical switching is a fundamental element for constructing all optical communication networks and all-optical computation. Traditional all-optical switching is based on the large nonlinearity in absorbing medium induced by strong control fields [6, 7], but it is difficult to satisfy the integrated manufacturing and compactness requirements due to its expensive cost and complex configuration. All-optical switching based on electromagnetically induced transparency (EIT) may be realized at lower control field intensities, but it usually only works at fixed absorbing frequencies of an optical medium under the two-photon resonance conditions [8]. Furthermore, with the rapid increase of data volume, all-optical computation and quantum teleportation inevitably require all-optical switches to work in parallel with multiple channels and broad bandwidth [9, 10]. Recently, a multi-channel all-optical switching scheme based on destructive quantum interference in multi-level excitation in a cavity quantum electrodynamics (CQED) system is proposed, but the switching efficiency (lower than 0.6 in some channels) and response speed remain to be improved [11]. Here we propose a high-efficiency, multi-channel and broadband all-optical switching scheme based on coherent perfect absorption (CPA) in an atom-cavity system. An ensemble of four-level atoms confined in a cavity are excited by a single cavity mode excited by two input fields injected from two ends of the cavity. Thus multiple output polariton channels are created. Output intensities of all channels are completely suppressed when the CPA criterion is satisfied [12-14] and the CQED system acts as a perfect light absorber. However, the output light intensity at certain channel approximates 1 when a weak control field is present. Thus the CQED system at certain channel acts as a perfect light reflector in this case. The proposed scheme can work in a broadband continuous frequency range and operate at weak light intensity with ultra-high efficiency. Such a CQED system has compactness and can be miniaturized, An all-optical cascade configuration based on multiple such CQED systems may be useful for practical all-optical switching and for applications such asall-optical modulators [15], all optical signal processors [16, 17], all-optical logic [18, 19], all-optical communications [20, 21], quantum information networks [22], etc.

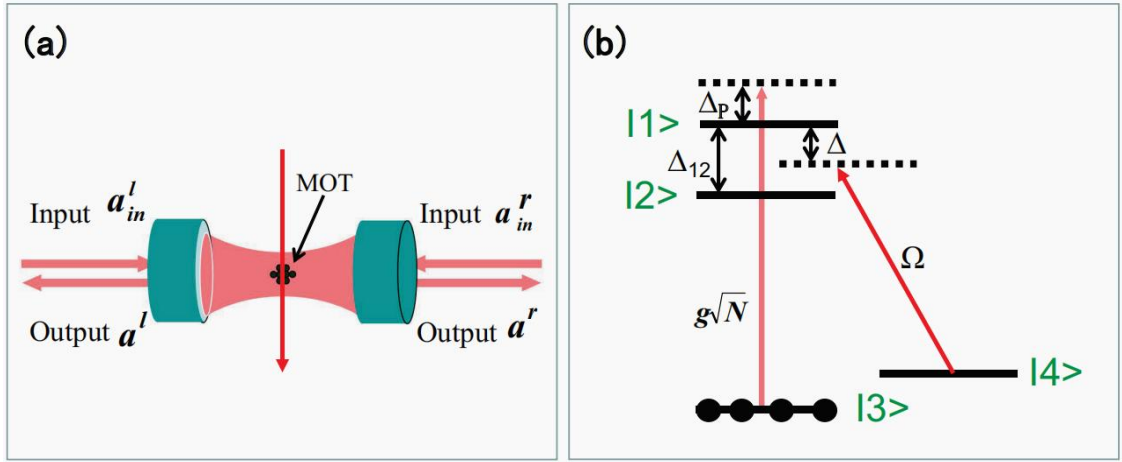

Fig. 1. Schematic diagrams of CQED setup and atomic energy-level configuration. (a) N four-level atoms are cooled in in a magneto-optical trap (MOT) and confined in a cavity. Two input fields are coupled into the cavity from two ends and a control field interacts with atomic ensemble from perpendicular direction in free space. (b) A single cavity mode excites two atomic transitions $|3\rangle \to |1\rangle$ and $|3\rangle \to |2\rangle$ simultaneously. A weak control field excites the atomic transition $|4\rangle \to |1\rangle$.

Fig. 1 shows the schematic diagram for the atom-cavity system. An ensemble of four-level atoms are cooled and trapped in a MOT, and confined ina single mode cavity [13, 23]. The cavity is excited by two input light fields $a_{in}^r$ and $a_{in}^l$ from two ends of the cavity. The input fields have identical frequency $\nu_p$ which is detuned from atomic transition $|3\rangle \to |1\rangle$ by $\Delta_p = \nu_p - \nu_{31}$. The optical cavity mode couples with two atomic transitions $|3\rangle \to |1\rangle$ and $|3\rangle \to |2\rangle$ simultaneously under the collective strong coupling condition [20, 24]. The cavity frequency detuning is defined as $\Delta_c = \nu_{cav} - \nu_{31}$. $\Delta_{12}$ denotes the frequency interval between two upper atomic energy levels $|1\rangle$ and $|2\rangle$. A control light only couples with atomic transition $|4\rangle \to |1\rangle$ and its frequency detuning is $\Delta = \nu_{con} - \nu_{41}$.

The interaction Hamiltonian of the CQED system is as following:

$$H = -\hbar\left(\sum_{i=1}^{N} g_1 \hat{a} \hat{\sigma}_{13}^{(i)} + \sum_{i=1}^{N} g_2 \hat{a} \hat{\sigma}_{23}^{(i)} + \sum_{i=1}^{N} g_1 \hat{a} \hat{\sigma}_{14}^{(i)} + i\hat{a}^+ \sqrt{2\kappa_1/\tau}\, a_{in}^r + + i\hat{a}^+ \sqrt{2\kappa_2/\tau}\, a_{in}^l \right) \quad (1)$$

here $\hat{a}$ and $\hat{a}^+$ are the annihilation and creation operators of intracavity photons respectively. The atom-cavity coupling coefficient $g_i = \mu_{j3}\sqrt{\omega_c/2\hbar\varepsilon_0 V}$ $(j=1,2)$ for the transition $|3\rangle \to |j\rangle$ is assumed to be uniform for N identical atoms inside the optical cavity. $\sigma_{mn}^{(i)}(m,n=1-4)$ denotes the atomic operator for the *i*th atom. Ω is the Rabi frequency of the control laser. Assume two cavity mirrors are identical and the loss rate of the cavity field on left and right mirrors are κ₁ and κ₂, respectively. κ₁ = κ₂ = κ. κ = T/τ. T is the mirror transmission rate and τ is the round trip time of photons inside the optical cavity. Under the rotating wave approximation [18], the energy nonconserving terms are dropped. According to the input-output theory of CQED [13], the intracavity light field a can be derived as:

$$\dot{a} = -\left[(\kappa_1+\kappa_2)/2 + i(\Delta_c - \Delta_p)\right]a + ig_1 N\sigma_{13} + ig_2 N\sigma_{23} + \sqrt{\kappa_1/\tau}\, a_{in}^r + \sqrt{\kappa_2/\tau}\, a_{in}^l \quad (2)$$

The system equations of the motion are solved by $\frac{d\hat{\rho}}{dt} = \frac{1}{i\hbar}[\hat{H},\hat{\rho}] + \hat{L}\hat{\rho}$ [25, 26]. $\hat{L}$ denotes the quantum superoperator. The CQED system is excited in linear regime when the intensity of signal light field $I_{in}$ is weak [11, 12]. The optical pumping effects for atoms from ground state to excited states can be neglected in this case, i.e., $N_3 \simeq 1$. The steady-state intracavity light field is derived as:

$$a = \frac{\sqrt{\kappa/\tau}\left(a_{in}^r + a_{in}^l\right)}{\kappa + i(\Delta_c - \Delta_p) - i\chi} \quad (3)$$

where $\chi = \frac{g_1^2 N(i+i\Omega\sigma_3)}{\gamma_{13}/2 - i(\Delta_p + \Delta_{12})} + \frac{ig_2^2 N}{\gamma_{23}/2 - i\Delta_p}$ is the atomic susceptibility. The denominator of the right side of Eq. (3) contains three imaginary terms when the control field is absent, so there are three transmission peaks in the cavity output spectrum. The transmission peak occurs when the system parameters meet the condition that the signal field phase shift from atoms and empty cavity cancel each other [27]. Assuming the collective-coupling coefficients of cavity mode to the two atomic transitions are identical, i.e., $g_j\sqrt{N} = g\sqrt{N}$, then the atomic susceptibility can be written as

$\chi = ig^2 N(d_1 + d_2 + d_2\Omega\sigma_3)$, where $\sigma_3 = \frac{-\Omega^* d_2}{\gamma_{43}/2 - i(\Delta - \Delta_p) + \Omega^2 d_2}$, $d_1 = \frac{2}{\gamma_{13} - 2i(\Delta_p + \Delta_{12})}$,

$d_2 = \dfrac{2}{\gamma_{23} - 2i\Delta_p}$. $\gamma_{mn}$ denotes the natural decay rate for atomic transition $|m\rangle \to |n\rangle$ and $\gamma_{13} = \gamma_{23} = \Gamma$. The output fields from two sides of the CQED system are derived as [13, 28]:

$$a^{r,l} = \dfrac{\kappa\left(a_{in}^r + a_{in}^l\right)}{\kappa + i(\Delta_c - \Delta_p) - i\chi} - a_{in}^{r,l} \qquad (4)$$

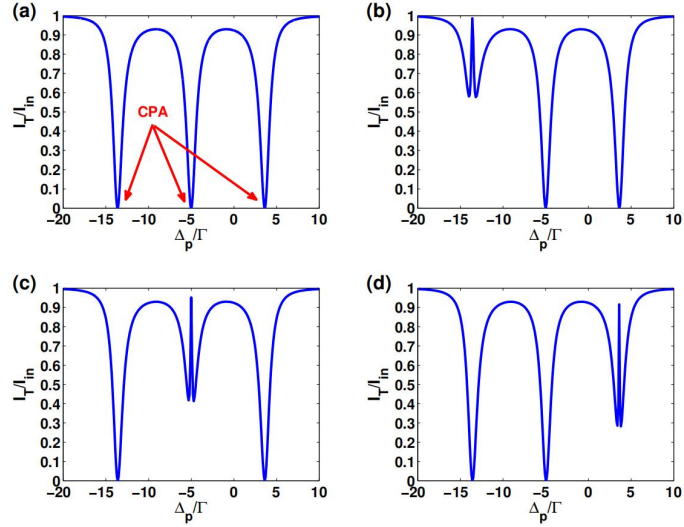

Fig. 2. The normalized output field intensity $I_T/I_{in}$ versus the signal frequency detuning $\Delta_p/\Gamma$. (a) The control light is absent ($\Omega = 0$); (b) The control light is presented and tuned to be resonant to the left polariton ($\Delta = -13.6\Gamma$, $\Omega = 0.5\Gamma$); (c) The control light is tuned to be resonant to the central polariton ($\Delta = -5\Gamma$, $\Omega = 0.5\Gamma$); (d) The control light is tuned to be resonant to the right polariton ($\Delta = 3.7\Gamma$, $\Omega = \Gamma$). Other parameters are $\Delta_c = -\dfrac{\Delta_{12}}{2}$, $\Delta_{12} = 10\Gamma$, $g\sqrt{N} = 5\Gamma$, and $\kappa = \Gamma$.

The intensities of the output signal field and the intracavity field are defined as $I_T = \left(a^{r,l}\right)^* a^{r,l}$ and $I = a^* a$ respectively. The input signal intensity is $I_{in} = (a_{in})^* a_{in}$. The normalized output (intracavity) field intensity is defined as $I_T/I_{in} (I/I_{in})$. Three polariton eigenstates are produced due to the collective strong coupling of the cavity mode with two separate atomic transitions [29]. The resonant frequencies of three polariton states depend on the cavity detuning $\Delta_c$, the collective coupling coefficient $g\sqrt{N}$, and the control light parameters. The excited polariton peaks are symmetrical when the cavity frequency is tuned to the middle of two excited atomic levels, i.e., $\Delta_c = -\dfrac{\Delta_{12}}{2}$ [11]. Two output fields at two ends of CQED system are equal when the input fields are identical, $a^{r,l} = a_{out}$

[12,13]. Without the contro light, CPA occurs when the CPA criterion $g\sqrt{N} = -\Delta_c$ is met (see Eq. (4), and then the output field is zero ($a_{out} = 0$),. Fig. 2 plots the output spectrum of the atom-cavity system. In Fig. 2(a), CPAs occur simultaneously at three polariton eigenstates when the CPA criterion is satisfied and without the control light present. A broadband multi-channel CPA can thus be realized by adjusting the cavity detuning $\Delta_c$, the collective coupling coefficient $g\sqrt{N}$, etc. The CQED system then acts as a broadband perfect absorber. Fig. 2(b-d) plot the all-optical switching operations when the control light is present and tuned to be resonant to the left polariton, central polariton and right polariton, respectively. A new narrow spectral peak is created at the resonant polariton eigenstate due to the destructive interference caused by the control light field with the multi-level excitation of the cavity mode, and the corresponding output light intensity of the system is changed from zero to maximum (approximate 1). However, CPAs at non-resonant polaritons still exist. Therefore, the output of the multi-channel signal fields can be turn on or off by the control light. Comparing to traditional all-optical switching which only works at a fixed frequency under the double resonance condition [30, 31], the bandwidth and operating channels of the proposed switching scheme are expanded in a frequency range that is greater than the normal-mode splitting separation ($2g\sqrt{N}$) of the CQED system [11, 24]. Fig. 3 plots the intracavity field intensity with and without the control light. Fig. 3(a) shows that the input light field energy are completely transferred to the intracavity excitation energy when the multi-channel CPAs occur. In Fig. 3(b-d), the EIT-like dip appears due to the destructive interference created by the control light. The intracavity light intensity is changed from maximum to zero, and the CQED system behave as a perfect lightreflector.

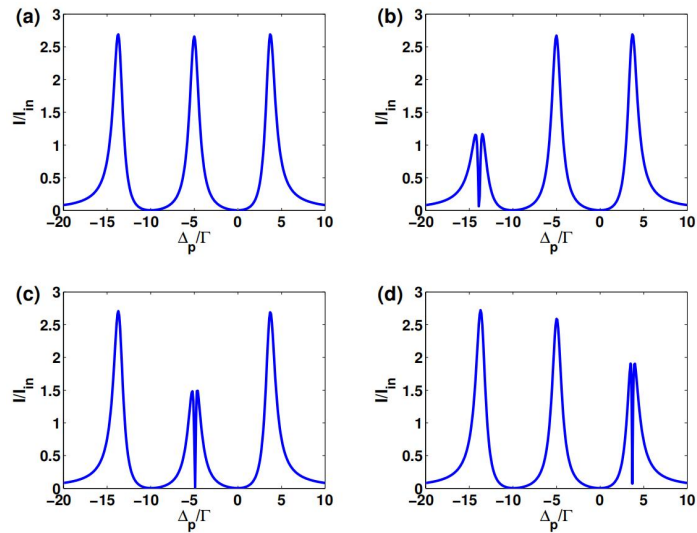

Fig. 3. The normalized intracavity field intensity $I/I_{in}$ versus the signal frequency detuning $\Delta_p/\Gamma$. (a) The control light is absent ($\Omega = 0$); (b) The control light is presented and tuned to be resonant to the left polariton ($\Delta = -13.9\Gamma$, $\Omega = 0.5\Gamma$); (c) The control light is tuned to be resonant to the central polariton ($\Delta = -5\Gamma$, $\Omega = 0.5\Gamma$); (d) The control light is tuned to be resonant to the right polariton ($\Delta = 3.8\Gamma$, $\Omega = \Gamma$). Other parameters are same as Fig. 2.

The efficiency of an optical switching can be defined as $\eta = \frac{I(|1\rangle) - I(|0\rangle)}{I(|1\rangle)}$. $I(|1\rangle)$ denotes the output signal intensity in the open state and $I(|0\rangle)$ denotes the output signal intensity in the closed state. For the proposed all-optical switching scheme, $I_T(\Omega \neq 0) = I_T(|1\rangle)$ denotes the closed state $|1\rangle$ while $I_T(\Omega = 0) = I_T(|0\rangle)$ denotes the open state $|0\rangle$. $\eta_T = \frac{I_T(\Omega \neq 0) - I_T(\Omega = 0)}{I_T(\Omega \neq 0)}$.

Similarly, the intracavity field can also be turned on or off by the control field. The intracavity field intensity reaches maximal value when the control field is absent, and vice versa. $\eta_I = \frac{I(\Omega = 0) - I(\Omega \neq 0)}{I(\Omega = 0)}$. Fig. 4 shows the switching efficiency of output signal $\eta_T$ and intracavity field $\eta_I$ versus the Rabi frequency $\Omega/\Gamma$ of the control field. The switching efficiency increases rapidly and saturates at $\Omega = 0.01\Gamma$ ($\eta = 1$). It is far below the saturation intensity of the control field transition and makes the single photon switching possible [5]. In Fig. 5, $\eta_T$ firstly increases and then decreases as $g\sqrt{N}$ increases. A maximal value $\eta_T \simeq 1$ is obtained simultaneously at multiple switching channels when $g\sqrt{N} = \Delta_c$, which is restricted by the CPA criterion and destructive interference effect. Compared to the optical switching scheme in Ref. [11], the proposed scheme based on CPA has ultra-high switching efficiency (approximate 1) at a broadband range, also, it calls much relaxed requirements for system parameters. Thus the proposed switching scheme is more robust and can be realized in many physical systems [20, 32]. To investigate the switching time, the frequency detuning $\Delta_p = \Delta \pm \delta$ of the signal laser is defined at which the output field intensity is $I(\delta) = (I_{max} + I_{min})/2$. $I_{max}$ and $I_{min}$ are the maximal value ($\Omega \neq 0$) and minimal value ($\Omega=0$) of output field intensity. The full widths at half maxima $2\delta$ in Fig. 2(b-d) are obtained as 1.4MHz, 1.3MHz and 1.22MHz. Correspondingly, the minimum switching times τ=1/4πδ correspondingly are 0.1μs, 0.12μs and 0.13μs, respectively.

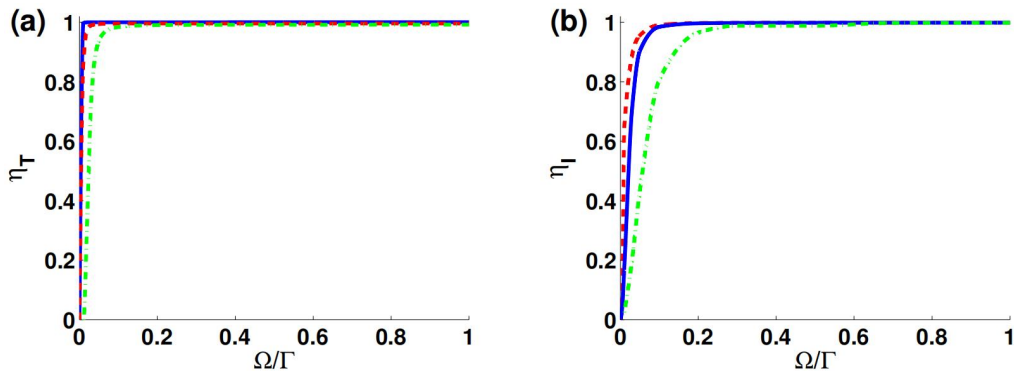

Fig. 4. Switching efficiency of (a) the output signal light $\eta_T$ and (b) the intracavity field $\eta_I$ versus $\Omega/\Gamma$. The red dashed, blue full and green dashed dotted lines correspond to the control field frequency resonant with the left, central and right polaritons. Other parameters are same as Fig. 2 and Fig. 3, respectively.

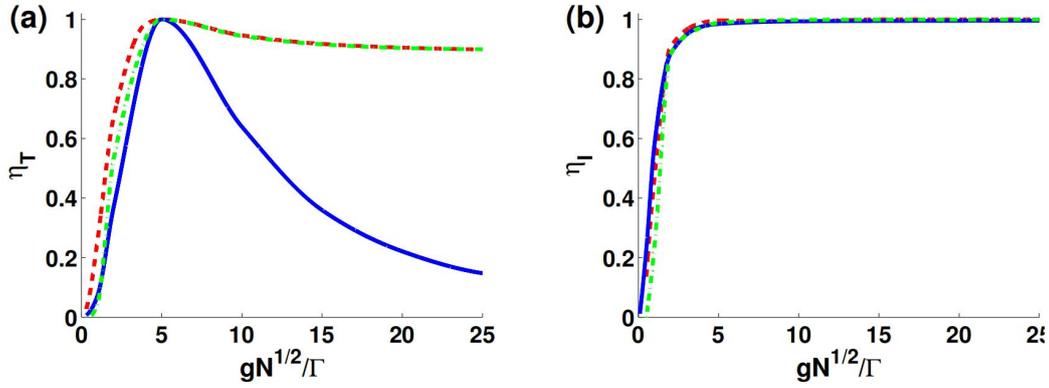

Fig. 5. Switching efficiency of (a) the output signal light $\eta_T$ and (b) the intracavity field $\eta_I$ versus $g\sqrt{N}/\Gamma$. The red dashed, blue full and green dashed dotted lines correspond to the control field frequency resonant with the left, central and right polaritons. Other parameters are same as Fig. 2 and Fig. 3, respectively.

The optical pumping effect cannot be neglected when the signal field intensity is strong, which renders the population of atoms in upper states to be nonzero, i.e., $N_1 \neq 0$, $N_2 \neq 0$. The CQED system is then driven into the nonlinear regime [11, 12]. Solving the nonlinear Schrödinger equation $i\hbar\frac{d\psi}{dt} = H\psi$ ($\psi = \sum_{i=1}^{4} c_i |\Phi_i\rangle$ $c_i$ is the probability amplitude of atom in state $|i\rangle$), the steady-state solutions of cavity input-output can be obtained. Similar to the case of linear regime, the CPA can be observed in the nonlinear regime of the CQED system [12, 26]. Fig. 6 shows the input-output properties of the CQED system with and without the control field. In Fig. 6(a), CPA occurs at central polariton excitation with $\Delta_p = \Delta_c$ and $\Omega=0$. Fig. 6(b) shows that the CPA is destroyed due to the destructive interference caused by the control light when the double resonance condition is satisfied ($\Delta = \Delta_p$). It shows that the control field can still be used to turn on or off the output of signal light in the nonlinear regime, and correspondingly the CQED system switches from a perfect reflector to a perfect absorber. However, unlike the monostability observed in the linear regime, the CQED system in the nonlinear regime exhibits bistable or even multistability behavior. In Fig. 6(c), the near CPA occurs when the signal field frequency is tuned to the middle between the left and central polaritons ($\Delta_p = -10\Gamma$) and the control field is absent.

In Fig. 6(d), the control laser is present and the double resonance condition is satisfied ($\Delta = \Delta_p$). Due to the destructive interference, the original near CPA disappears and a new CPA occurs at a lower threshold. The original bistable region is replaced by two new bistable regions with lower thresholds. Based on this, a broadband multi-throw all-optical switching can be designed which is widely used as basic device in all-optical communication network constructing [20, 21] and quantum teleportation [2]. Overall, the rich input-output characteristics of the CQED system in the nonlinear regime and associated bistablity and multistability behavior requires a more detailed analysis and will be a subject of a future study.

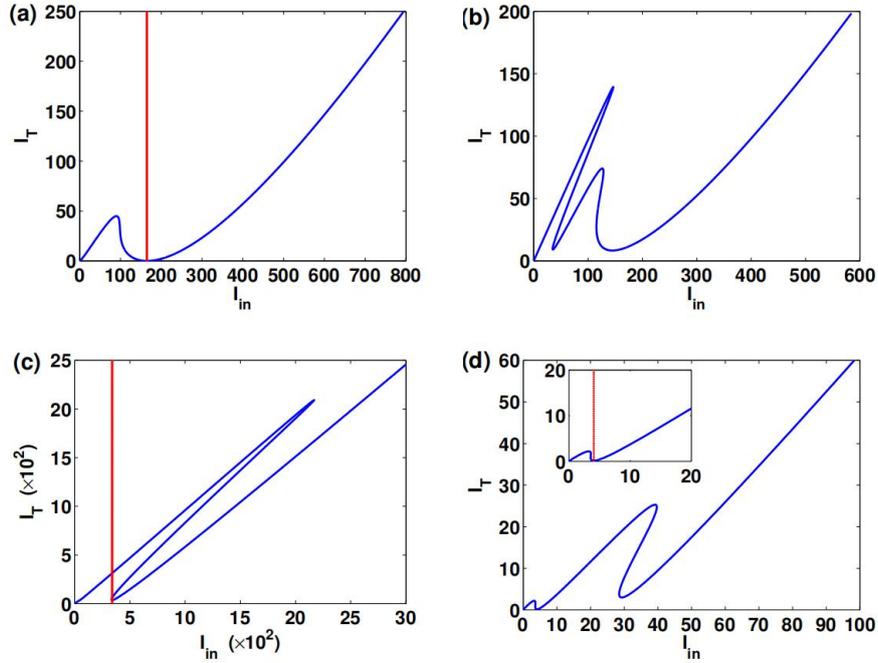

Fig. 6. The output field intensity $I_T$ versus the input field intensity $I_{in}$. The red lines label the thresholds of CPAs. (a) The control light is absent. $\Omega=0$, $\Delta_p = \Delta_c$; (b) The control light is present and tuned to be resonant to the central polariton. $\Delta = \Delta_p = \Delta_c$, $\Omega = 5\Gamma$; (c) The control light is absent. $\Omega=0$, $\Delta_p = 10\Gamma$; (d) The control light is present. $\Delta = \Delta_p = -10\Gamma$, $\Omega = 0.1\Gamma$. Inset plots the magnified input–output relation to show the new CPA and threshold clearly. Other parameters are $\Delta_c = -\dfrac{\Delta_{12}}{2}$, $\Delta_{12}=10\Gamma$, $g\sqrt{N}=2\sqrt{2}\Gamma$, and $\kappa=\Gamma$.

The proposed multi-channel all-optical switching scheme can be realized in many physical systems experimentally with moderate system parameter requirement [33-35]. For a practical example, a diode laser at 780 nm can be used a s the signal field to simultaneously couple the ground state $\left|^2S_{1/2}, F=1\right\rangle$ and two excited states $\left|^2P_{3/2}, F=0\right\rangle$ and $\left|^2P_{3/2}, F=1\right\rangle$ of [87]Rb D₂ transitions while Another diode laser at 780 nm can be used as the control field to couple the [87]Rb ground state $\left|^2S_{1/2}, F=2\right\rangle$ and excited state $\left|^2P_{1/2}, F=1\right\rangle$ transition.

To summarize, a multi-channel all-optical switching scheme based on CPA is proposed for a CQED system. The CQED system acts as a near perfect light absorber/reflector controlled by a weak control light field and works in both linear and nonlinear excitation regimes. In particular, a broadband multi-throw all-optical switching is demonstrated in nonlinear excitation regime. Compared to the traditional all-optical switching schemes which only work at certain fixed frequencies during they interact with absorbing media, the proposed all-optical switching scheme is tunable in frequency and has a wide frequency bandwidth and high switching speed. The proposed scheme is useful for applications such as single-photon detectors [37], all-optical routing [38, 39], all-optical communications [20, 21], all-optical quantum information processing and quantum logic elements [20, 22].

**Funding.** This work was supported by Wuhan University of Science and Technology, the Postdoctoral Applied Research Program of Qingdao (Grant No. 62350079311135), and Postdoctoral Applied Innovation Program of Shandong (Grant No. 62350070311227).

**Disclosures.** The authors declare no conflicts of interest.

**Data Availability Statement.** No data were generated or analyzed in the presented research.